\documentclass{emulateapj}
\usepackage{times,mathptm}
\shorttitle{
THE CHANDRA VISION OF HIGH-REDSHIFT QUASARS}
\shortauthors{VIGNALI ET AL.}

\newcommand{\ltsima}{$\; \buildrel < \over \sim \;$}
\newcommand{\simlt}{\lower.5ex\hbox{\ltsima}}
\newcommand{\gtsima}{$\; \buildrel > \over \sim \;$}
\newcommand{\simgt}{\lower.5ex\hbox{\gtsima}}

\newcommand{\lum}{\rm erg s$^{-1}$}

\def\sun{\hbox{$\odot$}}
\def\earth{\hbox{$\oplus$}}
\def\lesssim{\mathrel{\hbox{\rlap{\hbox{\lower4pt\hbox{$\sim$}}}\hbox{$<$}}}}
\def\gtrsim{\mathrel{\hbox{\rlap{\hbox{\lower4pt\hbox{$\sim$}}}\hbox{$>$}}}}

\def\arcdeg{\hbox{$^\circ$}}
\def\arcmin{\hbox{$^\prime$}}
\def\arcsec{\hbox{$^{\prime\prime}$}}

\def\aox{$\alpha_{\rm ox}$}

\def\lumh{\rm erg s$^{-1}$ Hz$^{-1}$}
\def\ab1450{$AB_{1450(1+z)}$}

\def\xray{\hbox{X-ray}}
\def\mb{$M_{\rm B}$}

\def\asca{{\it ASCA\/}}
\def\chandra{{\it Chandra\/}}

\def\heao1{{\it HEAO-1\/}}

\def\rosat{{\it ROSAT\/}}

\def\xmm{{XMM-{\it Newton\/}}}

\begin{document}

\title{X-ray Lighthouses of the High-Redshift Universe. II. \\
Further Snapshot Observations of the Most Luminous \boldmath$z\gtrsim4$ 
Quasars with \chandra}

\author{
C. Vignali,\altaffilmark{1,2} 
W.~N. Brandt,\altaffilmark{3} 
D.~P. Schneider,\altaffilmark{3} 
and S. Kaspi\altaffilmark{4,5}
}
\altaffiltext{1}{Dipartimento di Astronomia, Universit\`a degli Studi di 
Bologna, Via Ranzani 1, 40127 Bologna, Italy; cristian.vignali@bo.astro.it.}
\altaffiltext{2}{INAF--Osservatorio Astronomico di Bologna, Via Ranzani 1, 
40127 Bologna, Italy.}
\altaffiltext{3}{Department of Astronomy and Astrophysics, The Pennsylvania 
State University, 525 Davey Laboratory, University Park, PA 16802, USA; 
niel@astro.psu.edu \& dps@astro.psu.edu.}
\altaffiltext{4}{School of Physics and Astronomy, Raymond and Beverly Sackler 
Faculty of Exact Sciences, Tel-Aviv University, Tel-Aviv 69978, Israel; 
shai@wise.tau.ac.il.}
\altaffiltext{5}{Physics Department, Technion, Haifa, 32000, Israel.}

\begin{abstract}
We report on \chandra\ observations of a sample of 11 optically luminous 
(\mb$<-28.5$) quasars at \hbox{$z$=3.96--4.55} selected from the 
Palomar Digital Sky Survey and the Automatic Plate Measuring Facility 
Survey. These are among the most luminous $z\gtrsim4$ quasars 
known and hence represent ideal witnesses of the end of the ``dark age ''. 
Nine quasars are detected by \chandra, with $\approx$~2--57 counts 
in the observed \hbox{0.5--8~keV} band. 
These detections increase the number of \xray\ detected AGN at $z\gtrsim4$ 
to $\approx$~90; overall, \chandra\ has detected $\approx$~85\% of the 
high-redshift quasars observed with snapshot (few kilosecond) observations. 
PSS~1506$+$5220, one of the two \xray\ undetected quasars, displays a number 
of notable features in its rest-frame ultraviolet spectrum, the most 
prominent being broad, deep \ion{Si}{4} and \ion{C}{4} absorption lines. 
The average optical-to-X-ray spectral index for the present sample 
($\langle\alpha_{\rm ox}\rangle$=$-$1.88$\pm{0.05}$) is steeper than 
that typically found for $z\gtrsim4$ quasars but consistent with the 
expected value from the known dependence of this spectral index on 
quasar luminosity. 

We present joint \xray\ spectral fitting for a sample of 48 
radio-quiet quasars in the redshift range \hbox{3.99--6.28} 
for which \chandra\ observations are available. 
The \xray\ spectrum ($\approx$~870 counts) is 
well parameterized by a power law with $\Gamma$=1.93$^{+0.10}_{-0.09}$ 
in the rest-frame \hbox{$\approx$~2--40~keV} band, and a tight 
upper limit of $N_{\rm H}\approx$~5$\times10^{21}$~cm$^{-2}$ is 
obtained on any average intrinsic \xray\ absorption. 
There is no indication of any significant evolution in the \xray\ properties 
of quasars between redshifts zero and six, suggesting that the physical 
processes of accretion onto massive black holes have not changed over the 
bulk of cosmic time. 
\end{abstract}

\keywords{galaxies: active --- galaxies: nuclei --- quasars: general --- 
X-rays: galaxies}

\section{Introduction}
Over the last few years, \xray\ observations have provided important 
information about the population of luminous quasars populating the Universe 
at $z\approx$~4--6, i.e., at the end of the ``dark age'' (e.g., Rees 1999).
Since the work by Kaspi, Brandt, \& Schneider (2000) with 
archival \rosat\ observations, 
\xray\ studies of high-redshift quasars have taken advantage of 
the capabilities of the new generation of \xray\ observatories: 
\chandra\ and \xmm. 
In particular, \chandra\ snapshot observations of 
$z\gtrsim4$ Active Galactic Nuclei (AGN) 
have proven to be effective in providing basic \xray\ information, 
such as \xray\ fluxes, luminosities, and broad-band properties, 
on some of the most luminous objects in the Universe 
(Vignali et al. 2001a, 2002, 2003a,b, 
hereafter V01a, V02, V03a, and V03b, respectively; Brandt et al. 2002; 
Bechtold et al. 2003; Bassett et al. 2004). 
Using these observations, it has also been possible to collect enough source 
counts to allow joint \xray\ spectral 
fitting (V03a; V03b; Vignali, Brandt, \& Schneider 2004). 
This kind of \xray\ study is enabled by the sharp on-axis PSF 
and the low instrumental background of \chandra. 
On the other hand \xmm, with its larger effective area, 
has allowed individual \xray\ spectral fitting for a handful of the 
most \xray\ luminous $z>4$ quasars with medium-to-long exposures 
(Ferrero \& Brinkmann 2003; Grupe et al. 2004; Farrah et al. 2004; 
see also the \chandra\ spectral results of the $z=$5.99 
SDSS~130608.26$+$035626.3 quasar reported recently 
by Schwartz \& Virani 2004). 

It has therefore become possible to provide an overall picture of 
the radio-quiet and radio-loud quasar populations at the highest redshifts. 
However, \xray\ studies of the most optically luminous quasars 
at $z>4$ have been conducted for a limited number of objects, the majority 
of which have been selected from the Digital Palomar Sky Survey (PSS; e.g., 
Djorgovski et al. 1998) and presented by V03a. 
The goal of the present study is to extend previous \xray\ investigations 
of the properties of luminous quasars at $z\gtrsim4$ using \chandra\ snapshot 
observations; the target sample comprises 11 quasars in the redshift range 
\hbox{$z$=3.96--4.55} (\mb$\approx$~$-$28.5 to $-$29.5) selected from the PSS 
and the Automatic Plate Measuring Facility 
Survey (BRI; e.g., Irwin, McMahon, \& Hazard 1991; Storrie-Lombardi et 
al. 1996, 2001). 

Throughout the paper we adopt $H_{0}$=70 km s$^{-1}$ Mpc$^{-1}$, 
$\Omega_{\rm M}$=0.3, and $\Omega_{\Lambda}$=0.7 (e.g., Spergel et al. 2003).

\section{\chandra\ Observations and Data Reduction} 
All of the quasars presented here were targeted by \chandra\ 
during Cycle~4 with snapshot \hbox{($\approx$~4.0--5.1~ks)} observations. 
The observation log is shown in Table~1. 
All of the sources were observed with the Advanced CCD Imaging Spectrometer 
(ACIS; Garmire et al. 2003) with the S3 CCD at the aimpoint. 
Faint mode was used for the event telemetry format, and \asca\ grade 0, 2, 3, 
4 and 6 events were used in the analysis, which has been carried out using 
the \chandra\ Interactive Analysis of Observations ({\sc ciao}) Version 2.3 
software. No background flares are present in these observations. 
Source detection was carried out with {\sc wavdetect} (Freeman et al. 2002) 
similarly to V01a, V03a, and V03b, using 
wavelet transforms (with wavelet scale sizes of 1, 1.4, 2, 2.8, and 4 pixels) 
and a false-positive probability threshold of 10$^{-4}$. 
Given the small number of pixels being searched due to the known source 
positions and the sub-arcsec on-axis angular resolution of \chandra, 
the probability of spurious detections is extremely low; most of the sources 
were in fact detected at a false-positive probability threshold of 10$^{-6}$. 

The \xray\ counts detected in the ultra-soft band \hbox{(0.3--0.5~keV)}, 
the soft band \hbox{(0.5--2~keV)}, the hard band \hbox{(2--8~keV)}, and the 
full band \hbox{(0.5--8~keV)} are reported in Table~2. The counts have been 
derived using {\sc wavdetect} and checked with manual aperture photometry. 
Nine quasars have been detected, 
with full-band counts ranging between 2 and 57. 
The \xray\ positions of the detected sources 
are within 0.1--0.7\arcsec\ of the optical positions; 
this is consistent with the expected \chandra\ ACIS positional error. 

In Table~2 the band ratios and the effective photon indices for the 
\xray\ detected sources are also reported. Both of these quantities have been 
corrected for the quantum efficiency decay of ACIS at low energies, 
caused by molecular contamination of the ACIS filters, 
using a time-dependent correction. To apply the correction, we adopted the 
{\sc acisabs} model (see, e.g., Chartas et al. 
2002).\footnote{See http://www.astro.psu.edu/users/chartas/xcontdir/xcont.html.} 
We note that, given the dates of our observations, the differences 
between the adoption of either this model or that proposed by the 
\chandra\ \xray\ Center 
(CXC),\footnote{For further details, see http://asc.harvard.edu/ciao/why/acisqedeg.html.} available with the latest CIAO release, 
are small; therefore, for consistency with our previous work (e.g., V03a; 
V03b), we will adopt the {\sc acisabs} model throughout this paper. 

The full-band images of eight detected quasars, adaptively smoothed 
using the algorithm of Ebeling, White, \& Rangarajan (2005), 
are shown in Figure~1. 
For BR~1117$-$1329, a known broad absorption-line quasar (BALQSO; 
Storrie-Lombardi et al. 1996) from which two counts have been detected 
(see $\S$2.1 for discussion), and the two non-detections, 
the raw images are shown. 
Overall, in the regions of projected distance 
\hbox{$\approx650\times650$~kpc$^{2}$} centered on the quasar positions, 
we did not find any significant excess of companions with 
respect to the cumulative number counts from \xray\ surveys 
(e.g., Bauer et al. 2004); this is in agreement with previous 
\xray\ searches around high-redshift quasars (e.g., V03a; V03b). 
Similarly, the absence of \xray\ extension for all of our target quasars 
suggests that significant gravitational lensing 
(e.g., Comerford, Haiman, \& Schaye 2002) 
is not present.  

Although the observations are short, we have searched for \xray\ variability 
by analyzing the photon arrival times in the full band using the 
Kolmogorov-Smirnov test. 
No significant variability was detected from any of our quasars. This is not 
surprising given the small numbers of detected counts and the 
$\approx$~15-minute rest-frame exposure times.

\subsection{Two-Count Sources: BR~1117$-$1329 and PSS~1506$+$5220}
As was the case for the $z$=5.27 quasar SDSS~1208$+$0010 (V01a) and 
the $z$=4.80 quasar SDSS~0756$+$4502 (V03b), 
BR~1117$-$1329 can also be considered a secure 
detection with only two counts (in the same pixel) in the 
\hbox{0.5--2~keV band}. 
In fact, the source is detected by {\sc wavdetect} using a false-positive 
threshold of 10$^{-4}$. Monte-Carlo simulations 
(see $\S$2.1.1 of V03b for details) indicate that 
the significance of this detection is $\approx$~4$\sigma$ in the soft band. 

In contrast, PSS~1506$+$5220 cannot be considered detected (according to a 
Monte-Carlo analysis), although two full-band counts are present within the 
\hbox{2\arcsec--radius} circle centered on its optical position (see Fig.~1).

\section{Hobby-Eberly Telescope Observations}
We conducted imaging and spectroscopic observations of several 
of the quasars using the 9-m Hobby-Eberly Telescope (HET; 
Ramsey et al. 1998).  These data were taken within 1--10~months of the \xray\ 
observations \hbox{($ \le 8$ weeks} in the quasar rest frames) to minimize 
the effects of variability in the X-ray/optical comparisons.  
Variability can introduce significant uncertainties 
into estimates of key parameters (e.g., the optical-to-X-ray spectral index; 
see \S 4). The observations were all obtained with the Marcario Low-Resolution 
Spectrograph (LRS; Hill et al 1998a,b; Cobos Duenas et al. 1998).

\subsection{Photometric Observations}
Two-minute $R$-band LRS images were acquired for four of the 11 
sources (PSS~0747$+$4434, PSS~1506$+$5220, PSS~1646$+$5514, 
and PSS~2344$+$0342) 
during the fall--winter 2003--4.  Using \hbox{the $\approx 4' \times 4'$} 
LRS images and finding charts of the fields, it was possible to 
obtain approximate photometric calibrations using the stellar objects in 
the fields.  Although the level of uncertainty in these comparisons is 
substantial, we did not detect any significant long-term variations 
(i.e., more than 50\%) in the quasar brightnesses from their 
published values.  To have a uniform analysis, we have adopted the APM 
magnitudes\footnote{See http://www.ast.cam.ac.uk/$\sim$mike/apmcat/interface.html.} to derive the optical and broad-band parameters of the quasars.

\subsection{Optical Spectrum of PSS~1506$+$5220}
%
Two of the quasars, PSS~1058$+$1245 and PSS~1506$+$5220, do not have published 
spectra.  Immediately after acquiring the $R$-band image of PSS~1506$+$5220 
on January 19, 2004 (described in $\S$3.1), 
we obtained a 15-minute spectrum of the object using the LRS.  
The spectrum, which is displayed in Fig.~2, 
was acquired using a 2$''$ slit, an OG515 blocking filter, and a 
600 line mm$^{-1}$ grism; this configuration produced a spectrum covering the 
wavelength range 6300--9100~\AA\ at a resolving power of~$\approx$~1100.

The spectrum displays a number of notable and complex features, the 
most prominent being broad, deep \ion{Si}{4} and \ion{C}{4} absorption lines 
between redshifts of about 3.68 and 3.84 
(a velocity width of approximately 10,000~km~s$^{-1}$).  
The sharp rise at the blue end 
of the spectrum is the red wing of the Lyman~$\alpha$ emission line. 
Also present is a strong \ion{Mg}{2} absorption doublet (total rest equivalent 
width of nearly 7~\AA).

The previously reported redshift of PSS~1506$+$5220 is 
4.18;\footnote{See http://www.astro.caltech.edu/$\sim$george/z4.qsos.} 
however, accurate redshifts of $z>4$ BALQSOs are difficult to determine. 
In Fig.~2 we have indicated the expected position of the \ion{C}{4} emission 
line; no feature is found at this location.  
There is a suggestion of a broad emission 
feature centered at 7870~\AA; if this is the \ion{C}{4} line, the redshift 
would be approximately 4.08 (there is a hint of a narrow absorption feature 
in the midst of this putative emission line). 

G. Djorgovski has kindly showed an unpublished Keck spectrum of this object 
to us; his data extend far below the Lyman~$\alpha$ emission line.  If the 
peak of this line is associated with the rest wavelength of Lyman~$\alpha$,  
then the redshift is indeed 4.18.  Given the significant distortions in the 
lines produced by the absorption troughs, we conclude that it is 
presently impossible to assign a redshift with an accuracy of 
better than $\approx$~0.1 for this object. 
The ejection velocities of the BAL features would range between 
20,000 km~s$^{-1}$ and 25,000~km~s$^{-1}$, depending on the quasar redshift.
Throughout the analysis in this paper, we will adopt the published value
of 4.18 for the redshift of this object (see Footnote~9). 

The strong \ion{Mg}{2} absorption feature suggests the presence of a massive 
galaxy at a redshift of 1.4711 along the line-of-sight to the quasar. 
Given this result, the high luminosity of PSS~1506$+$5220 may partially 
arise from amplification via gravitational lensing.

\section{Multi-wavelength Properties of the Sample}
The principal optical, \xray, and radio properties of our target quasars 
are listed in Table~3: \\
{\sl Column (1)}. --- The name of the source. \\
{\sl Column (2)}. --- The Galactic column density (from Dickey \& 
Lockman 1990) in units of \hbox{10$^{20}$ cm$^{-2}$}. \\
{\sl Column (3)}. --- The monochromatic rest-frame \ab1450\ magnitude 
(\hbox{\ab1450=$-2.5\ \log f_{1450~\mbox{\scriptsize\AA}} -48.6$}; 
Oke \& Gunn 1983). 
The magnitudes reported in Table~3 have been derived from the APM $R$-band 
magnitudes (corrected for Galactic extinction; Schlegel, Finkbeiner, \& Davis 
1998) assuming the empirical relationship 
\hbox{\ab1450=$R-$0.684\ $z+3.10$}, 
which is effective in the redshift range covered by our sample. \\
{\sl Columns (4) and (5)}. --- The 2500~\AA\ rest-frame flux 
density and luminosity. These were computed from the \ab1450\ magnitude 
assuming an optical power-law slope of 
$\alpha=-0.79$ \hbox{($S_{\nu}$ $\propto$ $\nu^{\alpha}$}; Fan et al. 2001)  
to allow direct comparison with the results presented in V01a, V03a, and V03b. 
Note that the 2500~\AA\ rest-frame flux densities and luminosities are 
reduced by $\approx$~15\% when the power-law slope of the optical continuum 
is changed to $\alpha=-0.5$ (e.g., Schneider et al. 2001; 
Vanden Berk et al. 2001). \\
{\sl Column (6)}. --- The absolute $B$-band magnitude computed using 
\hbox{$\alpha=-0.79$}. When $\alpha=-0.5$ is adopted for the extrapolation, 
the absolute $B$-band magnitudes are fainter by $\approx$~0.35 mag. \\ 
{\sl Columns (7) and (8)}. --- The observed count rate in the 
\hbox{0.5--2~keV} band and the corresponding flux, corrected 
for Galactic absorption and the quantum efficiency decay of \chandra\ ACIS 
at low energy. 
The fluxes have been calculated using {\sc pimms} (Mukai 2002) and 
a power-law model with \hbox{$\Gamma$=2.0}; this is a good parameterization 
of samples of \hbox{$z\approx$~0--3} quasars (e.g., George et al. 2000; 
Reeves \& Turner 2000; Page et al. 2003; Piconcelli et al. 2005) 
and $z>4$ quasars (V03a; V03b; see $\S$6). \\
{\sl Columns (9) and (10)}. --- The rest-frame 2~keV flux density 
and luminosity, computed assuming $\Gamma$=2.0 and corrected for the quantum 
efficiency decay of \chandra\ ACIS at low energy. \\
{\sl Column (11)}. ---  The \hbox{2--10~keV} rest-frame luminosity. \\
{\sl Column (12)}. ---  The optical-to-X-ray power-law slope, \aox, 
defined as 
\begin{equation}
\alpha_{\rm ox}=\frac{\log(f_{\rm 2~keV}/f_{2500~\mbox{\rm \scriptsize\AA}})}{\log(\nu_{\rm 2~keV}/\nu_{2500~\mbox{\rm \scriptsize\AA}})}
\end{equation}
where $f_{\rm 2~keV}$ and $f_{2500~\mbox{\scriptsize \rm \AA}}$ are the 
rest-frame flux densities at 2~keV and 2500~\AA, respectively. 
The $\approx1\sigma$ errors on \aox\ have been computed following the 
``numerical method'' described in $\S$~1.7.3 of Lyons (1991) considering 
both the statistical uncertainties on the \xray\ count rates and the effects 
of reasonable changes in the \xray\ and optical continuum shapes 
(see $\S$3 of V01a for details). \\
{\sl Column (13)}. --- The radio-loudness parameter (e.g., Kellermann 
et al. 1989), defined as 
\hbox{$R$=$f_{\rm 5~GHz}/f_{\rm 4400~\mbox{\scriptsize\AA}}$} (rest frame). 
The rest-frame 5~GHz flux density was computed from the FIRST (Becker, White, 
\& Helfand 1995), NVSS (Condon et al. 1998) or Carilli et al. (2001) 
observed 1.4~GHz flux density assuming a radio power-law slope 
of $\alpha=-0.8$. The upper limits are at the 3\/$\sigma$ level. 
The rest-frame 4400~\AA\ flux density was computed from the \ab1450\ 
magnitude assuming an optical power-law slope of $\alpha=-0.79$. 
Typical radio-loudness values are $>100$ for radio-loud quasars (RLQs) 
and $<10$ for radio-quiet quasars (RQQs). 

For the two quasars with the lowest declinations, BR~0418$-$5723 and 
BR~2213$-$6729, no sensitive radio data are available. 
None of the remaining quasars is radio loud.

\section{X-ray Properties of the Sample}
In our previous study of a sample of optically luminous 
(\mb$\approx$~$-$28.4 to $-$30.2) $z>4$ PSS quasars (V03a), we derived the 
basic \xray\ properties for some of the most luminous objects in the 
Universe and performed joint \xray\ spectral fitting. 
No spectral evolution in quasar \xray\ continua at high redshift was found.  
With the current sample of 11 PSS and BRI quasars, our goal is to 
improve the results presented in V03a. 
The new \chandra\ observations, coupled with some other observations 
retrieved from the 
archive,\footnote{See http://www.astro.psu.edu/users/niel/papers/highz-xray-detected.dat 
for a regularly updated compilation of \xray\ detections and tight upper 
limits at $z\gtrsim4$. 
Details about the analyzed \chandra\ archival 
observations are reported in the Appendix.
have increased the number of \xray\ detected AGN at $z\gtrsim4$ 
(mostly quasars) to $\approx$~90. 
Overall, \chandra\ has detected $\approx$~85\% of the 
high-redshift quasars observed with snapshot observations. 
The program of snapshot observations has allowed filling of the soft 
\xray\ flux versus \ab1450\ plot shown in Fig.~3 with many \xray\ detections 
over the last few years. 
This plot is effective in showing the large advances made recently 
in the knowledge of the properties of high-redshift quasars 
(only RQQs are reported); for comparison, note the paucity of data points 
in Fig.~2 of Brandt et al. (2001). 
Most of the upper limits shown in Fig.~3 have been 
obtained using archival \rosat\ observations; some of these are reported here 
for the first time (see Appendix and Table~A1 for details).} 

Most of the quasars presented in this paper 
(filled triangles in Fig.~3 for the nine \xray\ detections and large 
downward-pointing arrows for the two upper limits) populate the 
region close to the \aox=$-$1.80 slanted line (computed assuming $z$=4.3, 
the average redshift of the current sample). 
They have an average $\langle\alpha_{\rm ox}\rangle$=$-$1.88$\pm{0.05}$ 
(computed using the 
{\sc asurv} software package Rev~1.2; LaValley, Isobe, \& Feigelson 1992; 
the quoted errors represent the standard deviation of the mean) 
which is steeper than the average value found using the 
whole population of RQQs at high redshift with available \xray\ information 
($\langle\alpha_{\rm ox}\rangle$=$-$1.76$\pm{0.02}$). 
%

This finding is consistent with the fact that the current \chandra\ sample is 
highly luminous \hbox{($\langle\log\ l_{2500}\rangle$=32.1~\lumh)}, 
and from previous work  (e.g., Avni, Worrall, \& Morgan 1995; 
Vignali, Brandt, \& Schneider 2003, hereafter VBS03; 
see also Strateva et al., in preparation) 
\aox\ has been found to depend upon 
optical luminosity, i.e., more optically luminous quasars have steeper 
(more negative) \aox. The expected \aox\ for the current sample would be 
$\approx$~$-1.8\pm{0.1}$. 
We note that \aox\ for BR~0353$-$3820 is flat 
(\aox=$-1.54^{+0.07}_{-0.06}$), much below the average value for the 
whole sample; it is possible that this flat \aox\ is due to some 
source variability occurring over the time interval between the 
optical and \xray\ observations. 

As discussed in previous works (e.g., V01a; V03b), there is a correlation 
between \ab1450\ and \hbox{0.5--2~keV} flux (see Fig.~3). 
To have a homogeneous sample, we have used only the $z\gtrsim4$ optically 
selected RQQs observed by \chandra; 
the significance of the correlation has been evaluated using the 
methods within {\sc asurv}, in particular the generalized 
Kendall's $\tau$ (Brown, Hollander, \& Korwar 1974). 
The \ab1450\ vs. \hbox{0.5--2~keV} flux correlation is significant at the 
$>99.9$\% level; according to the Estimate and Maximize 
(EM; Dempster, Laird, \& Rubin 1977) regression algorithm, it can be 
parameterized by 
\begin{eqnarray}
\nonumber
\log (F_{\rm X}/{\rm erg~cm^{-2}~s^{-1}})=[-(0.293\pm{0.057})\\ 
\nonumber
\times AB_{1450(1+z)}-(8.786\pm{1.059})], 
\end{eqnarray}
plotted as a dotted line in Fig.~3. 
%

Figure~3 also highlights the large discrepancy in our knowledge of the 
properties of the high-redshift AGN population in its entirety. 
While \xray\ information is presently available for large numbers 
of optically luminous quasars at $z\gtrsim4$ (i.e., those located on 
the right-hand side of Fig.~3 and plotted as filled symbols and 
downward-pointing arrows), only a few AGN at fainter optical 
magnitudes are detected in the \hbox{X-rays} (open squares). 
These have all been discovered by 
moderately deep and deep \xray\ surveys with \rosat\ 
(Schneider et al. 1998) and \chandra\ (Silverman et al. 2002; 
V02; Castander et al. 2003; Treister et al. 2004). 

Two quasars in the present sample have not been detected by \chandra: 
PSS~1506$+$5220 (discussed in $\S$2.1) 
and PSS~2344$+$0342 (no \xray\ counts within a \hbox{2\arcsec--radius} 
circle). 
There are two possible explanations for the non-detections of 
these two quasars in the \xray\ band: 
they are either strongly absorbed or intrinsically \xray\ weak. 
If their non-detection is due to the presence of intrinsic \xray\ absorption, 
then column densities of \hbox{$N_{\rm H}\simgt1.5\times10^{23}$~cm$^{-2}$} 
and \hbox{$N_{\rm H}\simgt3.3\times10^{23}$~cm$^{-2}$}, respectively, 
are required to reproduce the \chandra\ constraints (using 
the optical-to-X-ray correlation of VBS03 and assuming 
a photon index $\Gamma$=2.0). 
It is interesting to note that the HET spectrum of PSS~1506$+$5220 displays 
a number of notable and 
complex features, the most prominent being broad, deep \ion{Si}{4} and 
\ion{C}{4} absorption lines. 
For the other \xray\ undetected quasar of the present sample, 
PSS~2344$+$0342, the optical spectrum shows two 
very high column density damped Lyman~$\alpha$ absorption (DLA) systems 
(Peroux et al. 2001; Prochaska et al. 2003); however, these are unlikely to be 
responsible for all of the absorption suggested by the \xray\ constraints 
(see, e.g., Elvis et al. 1994 for discussion on the expected \xray\ 
absorption by DLAs). 

We also note that the \xray\ weak BR~1117$-$1329 (detected with two counts, 
see $\S$2.1 and Table~2) shows blueshifted broad absorption lines 
for \ion{O}{6}, \ion{N}{5}, \ion{Si}{4}, and \ion{C}{4} 
(Storrie-Lombardi et al. 1996).

\section{Joint Spectral Fitting: a Quasar X-ray Spectral Template 
at $z\approx$~4--6.3}
Previous studies have provided basic information on the \xray\ spectral 
properties of luminous, optically selected $z>4$ RQQs, 
in particular on their photon indices and column densities. 
These results have been achieved either with direct \xray\ 
spectral fitting of a few \xray\ luminous quasar spectra observed with \xmm\ 
(e.g., Ferrero \& Brinkmann 2003; Grupe et al. 2004; Farrah et al. 2004) 
and \chandra\ (Schwartz \& Virani 2004) 
or using joint \xray\ spectral fitting analysis, i.e., simultaneous 
fitting of sizable numbers of \chandra\ spectra of $z>4$ quasars 
(e.g., V03a; V03b; Vignali et al. 2004). 
Since these objects are among the most luminous quasars 
at $z>4$, they do not represent the majority of the high-redshift quasar population. 
Similarly to low- and intermediate-redshift RQQs (e.g., George et 
al. 2000; Reeves \& Turner 2000; Page et al. 2003; Piconcelli et al. 2005), 
high-redshift RQQs are, on average, characterized by 
power-law continua with \hbox{$\Gamma\approx$~1.8--2.0}, 
with no indication for widespread absorption. 
Joint \xray\ spectral fitting studies are 
important given the difficulty in obtaining quality \xray\ spectra of 
high-redshift quasars using reasonable exposures \hbox{($\approx$~40--60~ks)} 
with \xmm. 
Furthermore, combining the spectra of many high-redshift quasars is a 
practical way to derive their average 
\xray\ spectral properties; in fact, studies limited to a few \xray\ 
luminous quasars may not be representative of the high-redshift 
quasar population as a whole. 

To provide further constraints on the average \xray\ properties of 
high-redshift quasars, we have performed joint fitting using 
all the available \chandra\ data for RQQs 
in the redshift range \hbox{$z$=3.99--6.28}. 
Neither \xray\ selected quasars (see V02 for \xray\ spectral 
analyses of \xray\ selected $z>4$ AGN in the \chandra\ Deep Field-North) 
nor radio-selected quasars (with the exception of FIRST~0747$+$2739, 
which is radio quiet; see Bassett et al. 2004) 
have been considered in the following analysis. 
Source counts have been extracted for each quasar using a circular region 
of 2\arcsec\ radius centered on the \xray\ centroid of the quasar. 
Although background is typically negligible in these \chandra\ observations, 
we have extracted background counts using annuli of different sizes (to avoid 
contamination from nearby \xray\ sources) centered on the quasar position. 
Only sources with $>2$ full-band counts have been considered in the analysis. 
The selection criteria adopted in this study 
(mainly the RQQ selection and the fact that only \chandra\ data are used) 
reduced the number of $z\gtrsim4$ quasars with usable \xray\ 
information from $\approx$~90 to 48, for a total exposure time of 244.1~ks. 
We have $\approx$~870 net source counts in the observed \hbox{0.3--8~keV} 
band, corresponding to the rest-frame \hbox{$\approx$~1.5--58~keV} band; 
the average number of counts per object is $\approx$~18. 
Figure~4 summarizes the principal properties of the sample used in this 
\xray\ spectral study. The average redshift of the sample is $z$=4.57, 
while its median is $z$=4.43. 

Spectral analysis was carried out with {\sc xspec} Version 11.3.0 
(Arnaud 1996) using unbinned data and the {\it C}-statistic (Cash 1979); 
this statistical approach allows one to retain all 
spectral information and associate with each quasar 
its own Galactic absorption column density and redshift 
(for the fitting with intrinsic absorption, see below). 
Errors are quoted at the 90\% confidence level for one 
interesting parameter (\hbox{$\Delta$$C=2.71$}; Avni 1976; Cash 1979), 
unless stated otherwise. 
Solar abundances (from Anders \& Grevesse 1989) have been adopted 
for the absorbing material, although optical studies indicate 
that high-redshift quasars 
are often characterized by supersolar abundances of heavy elements 
(e.g., Hamann \& Ferland 1999; Dietrich et al. 2003a,b). 
Note that doubling the abundances in the fit produces 
a reduction in the column density of a factor of $\approx$~2. 

We started the joint \xray\ spectral fitting using a power-law model, leaving 
the power-law normalizations for all of the quasars free to vary 
(to account for the different fluxes of the sources), 
plus Galactic absorption. 
The lack of significant data-to-model residuals for the fit 
suggests that the assumed model is acceptable; 
the resulting photon index, $\Gamma$=1.93$\pm{0.09}$, is 
consistent with previous estimates for $z>4$ quasars. 
The addition of neutral intrinsic absorption at each source's redshift does 
not improve the fit significantly; there is no evidence for 
widespread absorption toward the lines-of-sight of the quasars 
under investigation, the column density (i.e., the counts-weighted average 
column density) being $\simlt5.0\times10^{21}$~cm$^{-2}$ 
(see the solid contours 
in Fig.~5a showing the 68, 90, and 99\% confidence levels for the photon 
index vs. column density; see also Table~4). 
If our targets had intrinsic absorption of 
\hbox{$N_{\rm H}\approx$~(2--5)$\times10^{22}$~cm$^{-2}$}, as found in two 
RQQs at $z\approx$~2 using \asca\ data (Reeves \& Turner 2000), we would 
be able to detect it given the upper limit on the column density we derived. 
The spectral results do not change significantly if only the \xray\ data 
in the rest-frame interval common to all of the quasars of the sample 
(\hbox{$\approx$~2.2--40~keV} band; $\approx$~840 counts) 
are taken into account (see the dashed contours in Fig.~5a). 
In this case, the photon index is $1.95\pm{0.10}$, 
while the upper limit on the column density remains essentially unchanged. 

The sample does not appear to be biased by the presence of a few high 
signal-to-noise ratio objects.\footnote{For this reason, we have 
considered the $\approx$~8.1~ks \chandra\ observation of 
SDSS~130608.26$+$035626.3 (Brandt et al. 2002) instead of the 
longer ($\approx$~120~ks) observation presented recently by 
Schwartz \& Virani (2004).} 
This has been checked during the \xray\ 
spectral analysis by removing some of the quasars with higher photon 
counting statistics. In particular, the removal of the three sources 
with $\gtrsim$60 counts (BR~0353$-$3820, PSS~0926$+$3055, 
and PSS~1326$+$0743) leaves the photon index unchanged 
($\Gamma$=1.92$\pm{0.10}$), 
but the constraint on the column density intrinsic to the quasars is 
less tight ($\simlt7.5\times10^{21}$~cm$^{-2}$; see Table~4). 
We also performed joint spectral fitting using the most and least 
optically luminous RQQs of the sample as well as the highest and lowest 
redshift RQQs; the results of the fitting are reported in Table~4. 
Within the statistical uncertainties due to the limited number of \xray\ 
counts, no evidence for a significant dependence of the \xray\ spectral 
parameters upon either the luminosity or the redshift of the quasar 
subsample has been found.

Given the limited counting statistics available 
and the fact that a power-law model seems to be acceptable, 
we did not investigate more complex spectral models. 
The typical upper limits on the equivalent width (EW) of a narrow, 
either neutral or ionized, iron K$\alpha$ line are not stringent, 
lying in the range \hbox{$\approx$~270--370~eV}. 
Using the parameterization of the anti-correlation between the EW of the 
iron K$\alpha$ line and source \xray\ luminosity (the ``\xray\ Baldwin 
effect''; see Iwasawa \& Taniguchi 1993 and Nandra et al. 1997) 
recently reported by Page et al. (2004), 
we would expect EW$\approx$~50~eV for our sample of RQQs 
(their average rest-frame 2--10~keV luminosity is $10^{45.3}$~\lum). 
The expected EW is below the upper limits provided by joint \xray\ 
spectral fitting. 

To investigate the reliability of the joint \xray\ spectral fitting results, 
we performed two consistency checks. 
First, we compared the photon index obtained from the joint spectral fitting 
with that derived from band-ratio analysis 
(i.e., the ratio of the \hbox{2--8~keV} to \hbox{0.5--2~keV} counts, 
corrected for the ACIS quantum efficiency decay); for the Galactic column 
density, we assumed the average value for the 48 quasars weighted by the 
exposure time of each observation. 
The resulting photon index, $\Gamma$=1.89$\pm{0.08}$ (the uncertainty quoted 
here corresponds to the 1$\sigma$ level, see Gehrels 1986 and the 
caption of Table~2), is consistent with our previous value. 
Secondly, we added all of the source and background counts using the 
{\sc ftool} task {\sc mathpha}; similarly, redistribution matrix files (RMFs) 
and auxiliary response files (ARFs) were combined using {\sc addrmf} 
and {\sc addarf}, respectively, weighted by the individual exposure times. 
In this case the source counts were grouped 
into a spectrum such that each spectral bin contained at least 15 counts to 
allow $\chi^{2}$ fitting. Our fitting obtained $\Gamma$=1.97$\pm{0.11}$ and no 
evidence for intrinsic absorption 
\hbox{($N_{\rm H}\simlt4.4\times10^{21}$~cm$^{-2}$)}, 
as for the previously adopted fitting procedures. 
The power-law model and the data-to-model residuals are shown in Fig.~5b. 

Fig.~6a shows the hard photon index (to avoid spectral complexities 
such as soft excesses at low energies) as a function of redshift for 
a compilation of optically selected RQQs. 
In particular, at low ($z<1.8$) redshift we have used the \xray\ spectral 
information for the Bright Quasar Survey (PG; Schmidt \& Green 1983) objects 
presented by Piconcelli et al. (2005; circles); 
a few sources with Seyfert-like luminosity (\mb$>-23$) have been excluded 
from this analysis. 
At $z\approx1.8-2.5$, the RQQs from the Vignali et al. (2001) sample 
have been selected (diamonds). 
At the highest redshifts, the \xray\ spectral results obtained over the 
last three years with either joint spectral fitting 
(triangles) or individual observations (squares) are shown. 
The low-redshift ($z\simlt1$) quasars plotted in Fig.~6a are 
usually $\approx$~4 magnitudes less luminous than the $z\gtrsim4$ quasars 
under study; we are able to get down only to $z\approx1$ with comparably 
luminous objects. 
Although significant intrinsic scatter in the photon indices at 
all redshifts is present, there is no detectable systematic change in 
the photon-index distribution at high redshift.
The apparently smaller dispersion in the photon indices at 
high redshift than at lower redshift is at least partially due to the 
joint spectral fitting (used for some data points at $z>4$) which tends to 
average any dispersion out. 
Although it is possible in principle that a redshift-dependence of 
$\Gamma$ could be canceled out by a luminosity-dependence of $\Gamma$ 
(which is not evident from Fig.~6b, where $\Gamma$ is plotted against \mb), 
this would require somewhat of a conspiracy and cuts agains Occam's razor. 
At present there is no overwhelming evidence that the photon index 
is significantly dependent upon quasar \xray\ luminosity; for further details, 
see the contrasting results presented by Dai et al. (2004) 
and Reeves \& Turner (2000). 
A partial correlation analysis of the photon index vs. 
redshift/\xray\ luminosity for a large, uniformly selected sample of quasars 
is required to address this issue properly. 

All of these spectral results confirm the lack of significant \xray\ spectral 
evolution of luminous, optically selected quasars over cosmic time. 
Thus, despite the strong changes in environment and quasar number 
density that have occurred from \hbox{$z\approx$~0--6}, individual quasar 
\xray\ emission regions seem to evolve relatively little 
(see Brandt et al. 2005).

\section{X-ray undetected quasars: peculiar objects at high redshift?}
Although the \xray\ spectral results presented above are based on a 
significant fraction of the RQQs with \chandra\ observations, 
the selection of the sample might constitute a cause for concern, 
since we have excluded the \xray\ faintest (with less than two counts) sources 
and non-detections. 
Overall, these \xray\ weak or undetected quasars represent $\approx$~25\% of 
the RQQ population thus far observed by \chandra\ at $z\gtrsim4$. 
It is possible that in some cases the lack of an \xray\ detection is due 
to absorption, which should be $\gtrsim$~10$^{23}$~cm$^{-2}$ to reproduce 
the \xray\ constraints (under reasonable assumptions about the spectrum). 
This is an acceptable explanation for BALQSOs (40\% of the the \xray\ 
undetected quasars, see Fig.~7). Past works have shown that BALQSOs are 
typically absorbed in the \xray\ band 
(e.g., Brandt et al. 2000; Green et al. 2001; V01a; Gallagher et al. 2002). 

In the \aox\ histogram shown in Fig.~7, BALQSOs populate preferentially 
the fainter part (i.e., more negative \aox values) of the distribution. 
Although computed using a limited number of objects, their average \aox\ 
(dotted line in Fig.~7) is $-1.96\pm{0.06}$ ($-1.92\pm{0.06}$ including 
SDSS~104433.04$-$012502.2 detected by \xmm; see Brandt et al. 2001), 
while the optically selected non-BAL RQQs observed with \chandra\ have 
$-1.76\pm{0.02}$ (long-dashed line in Fig.~7; the median value is $-$1.77). 
It is also plausible, though not necessarily required by the present data, 
that some of the remaining \xray\ undetected quasars at 
high redshift are characterized by absorption features which are not 
recognized because of either low spectral resolution (see, e.g., the case of 
SDSS~173744.87$+$582829.5 in V03b, where narrow absorption features were found 
with the HET) or limited spectral coverage (e.g., Maiolino et al. 2001, 
2004; Goodrich et al. 2001). 

Given this, we believe that the criteria adopted in the selection of the 
high-redshift non-BAL RQQs for \xray\ spectral analysis (see $\S$6) 
are not a cause for concern and the results we derived can be considered 
representative of the RQQ population at $z\gtrsim4$.

\section{Summary}
We have analyzed \chandra\ observations of a sample of 11 quasars at 
\hbox{$z\approx$~3.96--4.55} selected from the PSS and APM surveys. 
These are among the most luminous $z\gtrsim4$ quasars known and 
represent ideal probes of the dawn of the modern Universe. 
Given the optical luminosities of these quasars 
(\mb$\approx$~$-28.5$ to $-$29.5), the present study can be considered 
complementary to past work (e.g., V03a; V03b). 
The principal results of this work are \\
\begin{enumerate}
\item 
Nine of the quasars have been detected, thus increasing the number of \xray\ 
detected AGN at $z\gtrsim4$ to $\approx$~90. 
Including also the two \xray\ non-detections, 
the average \aox\ of the present sample, $-1.88\pm{0.05}$, is steeper 
that typically found for $z\gtrsim4$ quasars, but consistent with the value 
expected applying the known dependence of this spectral index upon 
quasar luminosity. 
\item
Two quasars are not detected, PSS~1506$+$5220 and PSS~2344$+$0342. 
If absorption intrinsic to the source were the cause of the \xray\ weakness 
of these two objects, then column densities larger than 
\hbox{$N_{\rm H}\approx1.5\times10^{23}$~cm$^{-2}$} would be implied. 
In the case of PSS~1506$+$5220, 
we found evidence for notable features, 
the most prominent being broad absorption lines from 
\ion{Si}{4} and \ion{C}{4}. For this source 
it is possible that the UV absorber is linked to the absorption 
likely present in the \xray\ band. This has been observed before in 
other BALQSOs at high redshift. 
\item
Using all the available RQQs at $z\gtrsim4$ detected by \chandra\ with 
more than 2 counts, we have derived \xray\ spectral constraints for the 
quasar population at high redshift. A power law with 
$\Gamma$=1.93$\pm{0.09}$ is a good parameterization of the quasar 
spectra in the rest-frame \hbox{$\approx$~2--40~keV band}, similar 
to what is found for the hard \xray\ continua of 
low- and intermediate-redshift quasars. 
Our results are consistent with a lack of significant spectral evolution of 
quasar \xray\ continua over cosmic time. A partial correlation 
analysis of the photon index vs. redshift/luminosity for a uniform 
sample of quasars over a broad range of redshift and luminosity is required 
to address this issue further.

The tight constraint on the absorption intrinsic to the quasars, 
$N_{\rm H}\simlt5.0\times10^{21}$~cm$^{-2}$, indicates the lack of widespread 
obscuration. 
\end{enumerate}

\acknowledgments
The authors thank G. Djorgovski for kindly sharing data, 
G. Chartas and C. Peroux 
for useful discussions, L. Angeretti for help in producing the figures, and 
M. Mignoli for help with {\sc IRAF} tasks. 
The authors thank the referee for his/her constructive comments. 
CV acknowledges partial support from MIUR (COFIN grant 03-02-23). 
The financial support of NSF CAREER award AST-9983783 
(WNB), CXC grant GO3-4143X (WNB), and NSF grant AST03-07582 (DPS) 
is also acknowledged. 

The HET is a joint project of the University of Texas 
at Austin, the Pennsylvania State University, Stanford University, 
Ludwig-Maximillians-Universit\"at M\"unchen, 
and Georg-August-Universit\"at G\"ottingen. 
The HET is named in honor of its principal benefactors, 
William P. Hobby and Robert E. Eberly. The Marcario Low-Resolution 
Spectrograph is named for Mike Marcario of High Lonesome Optics, who 
fabricated several optics for the instrument but died before its completion; 
it is a joint project of the HET partnership and the 
Instituto de Astronom\'{\i}a de la Universidad Nacional Aut\'onoma de M\'exico.

\clearpage

\begin{deluxetable}{lccccccc}
\tablecolumns{8} 
\tabletypesize{\footnotesize}
\tablewidth{0pc} 
\tablecaption{\chandra\ Observations of High-Redshift Quasars}
\tablehead{
\colhead{Object} & \colhead{} & \colhead{Optical} & \colhead{Optical} & 
\colhead{$\Delta_{\rm Opt-X}$\tablenotemark{a}} & 
\colhead{\xray} & \colhead{Exp.~Time\tablenotemark{b}} & \colhead{} \\
\colhead{Name} & \colhead{$z$} & \colhead{($\alpha_{2000}$)} & 
\colhead{($\delta_{2000}$)} & \colhead{(arcsec)} & \colhead{Obs. Date} & 
\colhead{(ks)} & \colhead{Ref.} 
}
\startdata 
BR~0331$-$1622    & 4.36 & 03 34 13.4 & $-$16 12 04.8 & 0.4     & 2003 Jun 17   & 4.67 & 1,2 \\
BR~0353$-$3820    & 4.55 & 03 55 04.9 & $-$38 11 42.3 & 0.3     & 2003 Sep 10   & 4.06 & 1,2 \\
BR~0418$-$5723    & 4.46 & 04 19 50.9 & $-$57 16 13.1 & 0.7     & 2003 Jun 15   & 3.99 & 1,2 \\
BR~0424$-$2209    & 4.32 & 04 26 10.3 & $-$22 02 17.6 & 0.4     & 2002 Dec 14   & 4.67 & 1,2 \\
PSS~0747$+$4434   & 4.43 & 07 47 49.7 & $+$44 34 16.9 & 0.5     & 2002 Dec 17   & 4.54 & 1 \\	
PSS~1058$+$1245   & 4.33 & 10 58 58.4 & $+$12 45 54.8 & 0.1     & 2003 Mar 2--3 & 5.05 & 3 \\
BR~1117$-$1329    & 3.96 & 11 20 10.3 & $-$13 46 25.7 & 0.4     & 2003 Jan 28   & 4.70 & 4 \\
PSS~1506$+$5220   & 4.18 & 15 06 54.5 & $+$52 20 05.2 & \nodata & 2003 Mar 14   & 4.86 & 3 \\	
PSS~1646$+$5514   & 4.04 & 16 46 56.4 & $+$55 14 46.4 & 0.3     & 2003 Sep 10   & 4.85 & 1 \\	
BR~2213$-$6729    & 4.47 & 22 16 51.9 & $-$67 14 43.4 & 0.4     & 2003 Mar 10   & 4.90 & 1,2 \\
PSS~2344$+$0342   & 4.24 & 23 44 03.2 & $+$03 42 25.5 & \nodata & 2003 Nov 20   & 5.10 & 1 \\	
\tableline
\enddata
\label{tab1}
\tablecomments{The optical positions of the quasars have been obtained using 
{\sc SExtractor} (Bertin \& Arnouts 1996) on the DPOSS2 images, 
while the \xray\ positions have been obtained with {\sc wavdetect}.}
\tablenotetext{a}{Distance between the optical and \xray\ positions; 
the dots indicate no \xray\ detection.}
\tablenotetext{b}{The \chandra\ exposure time has been corrected for detector 
dead time.}
\tablerefs{
(1) Peroux et al. 2001; (2) Storrie-Lombardi et al. 2001; 
(3) Omont et al. 2001; (4) Storrie-Lombardi et al. 1996.}
\label{tab1}
\end{deluxetable}

\clearpage

\begin{deluxetable}{lcccccc}
\tablecolumns{7}
\tabletypesize{\footnotesize}
\tablewidth{0pt}
\tablecaption{
X-ray Counts, Band Ratios, and Effective Photon Indices}
\tablehead{ 
\colhead{} & \multicolumn{4}{c}{X-ray Counts$^{\rm a}$} \\
\cline{2-5} \\
\colhead{Object} & \colhead{[0.3--0.5~keV]} & \colhead{[0.5--2~keV]} & 
\colhead{[2--8~keV]} & \colhead{[0.5--8~keV]} & 
\colhead{Band Ratio\tablenotemark{b}} & \colhead{$\Gamma$\tablenotemark{b}}
}
\startdata
BR~0331$-$1622    & $<3.0$              & {\phn}8.9$^{+4.1}_{-2.9}$ & {\phn}3.9$^{+3.2}_{-1.9}$ & 12.6$^{+4.7}_{-3.5}$ 
& 0.34$^{+0.31}_{-0.19}$ & 1.6$^{+0.7}_{-0.6}$ \\
BR~0353$-$3820    & 4.0$^{+3.2}_{-1.9}$ & 45.9$^{+7.8}_{-6.8}$      & 10.9$^{+4.4}_{-3.3}$      & 56.7$^{+8.6}_{-7.5}$ 
& 0.18$^{+0.08}_{-0.06}$ & 2.0$\pm{0.3}$ \\
BR~0418$-$5723    & $<3.0$              & {\phn}7.0$^{+3.8}_{-2.6}$ & $<6.4$                    & {\phn}9.0$^{+4.1}_{-2.9}$ 
& $<0.70$ & $>0.8$ \\
BR~0424$-$2209    & $<4.8$              & 10.0$^{+4.3}_{-3.1}$      & $<6.4$                    & 11.9$^{+4.6}_{-3.4}$ 
& $<0.49$ & $>1.2$ \\
PSS~0747$+$4434   & $<4.8$              & {\phn}6.0$^{+3.6}_{-2.4}$ & $<3.0$                    & {\phn}6.0$^{+3.6}_{-2.4}$ 
& $<0.39$ & $>1.6$ \\
PSS~1058$+$1245   & $<4.8$              & {\phn}3.0$^{+2.9}_{-1.6}$ & $<6.4$                    & {\phn}5.0$^{+3.4}_{-2.2}$ 
& $<1.64$ & $>0.1$ \\
BR~1117$-$1329    & $<3.0$              & {\phn}2.0$^{+2.7}_{-1.3}$ & $<3.0$                    & {\phn}2.0$^{+2.7}_{-1.3}$ 
& $<1.15$ & $>0.5$ \\
PSS~1506$+$5220   & $<3.0$              & $<4.8$                    & $<4.8$                    & $<6.4$ 
& \nodata & \nodata \\
PSS~1646$+$5514   & $<3.0$              & {\phn}6.0$^{+3.6}_{-2.4}$ & $<3.0$                    & {\phn}6.0$^{+3.6}_{-2.4}$ 
& $<0.38$ & $>1.4$ \\
BR~2213$-$6729    & $<3.0$              & 17.9$^{+5.3}_{-4.2}$      & {\phn}3.0$^{+2.9}_{-1.6}$ & 20.9$^{+5.7}_{-4.5}$ 
& 0.13$^{+0.13}_{-0.07}$ & 2.3$^{+0.7}_{-0.6}$ \\
PSS~2344$+$0342   & $<3.0$              & $<3.0$                    & $<3.0$                    & $<3.0$ 
& \nodata & \nodata \\
\tableline
\enddata
\tablenotetext{a}{
Errors on the \xray\ counts were computed according to Tables~1 and 2 
of Gehrels (1986) and correspond to the 1$\sigma$ level; these were 
calculated using Poisson statistics.  The upper limits are at the 95\% 
confidence level and were computed according to Kraft, Burrows, \& 
Nousek (1991).  For sake of clarity, upper limits of 3.0, 4.8, and 6.4 
indicate that 0, 1, and 2 \xray\ counts, respectively, have been found 
within an extraction region of radius 2\arcsec\ centered on the optical 
position of the quasar (considering the background within this source 
extraction region negligible).}
\tablenotetext{b}{
We calculated errors at the $\approx1\sigma$ level for the band ratio and 
effective photon index following the ``numerical method'' 
described in $\S$~1.7.3 of Lyons (1991); this avoids the failure of the 
standard approximate variance formula when the number of counts is small 
(see $\S$~2.4.5 of Eadie et al. 1971). 
The band ratios and photon indices have been obtained applying the correction 
required to account for the ACIS quantum efficiency decay at low energy.}
\label{tab2}
\end{deluxetable}

\clearpage

\begin{turnpage}
\begin{deluxetable}{ccccccccccccc}
\tablecolumns{13}
\tabletypesize{\footnotesize}
\tablewidth{0pt}
\tablecaption{Optical, X-ray, and Radio Properties of the Sample of High-Redshift Quasars}
\tablehead{
\colhead{Object} & \colhead{$N_{\rm H}$\tablenotemark{a}} & 
\colhead{$AB_{1450(1+z)}$} & \colhead{$f_{2500}$\tablenotemark{b}} & \colhead{$\log (\nu L_\nu )_{2500}$} & 
\colhead{$M_B$} & \colhead{Count~rate\tablenotemark{c}} & 
\colhead{$f_{\rm x}$\tablenotemark{d}} & \colhead{$f_{\rm 2\ keV}$\tablenotemark{e}} & 
\colhead{$\log (\nu L_\nu )_{\rm 2\ keV}$} & \colhead{$\log (L_{\rm 2-10~keV})$\tablenotemark{f}} & 
\colhead{$\alpha_{\rm ox}$\tablenotemark{g}} & \colhead{$R$\tablenotemark{h}} \\
\colhead{(1)} & \colhead{(2)} & \colhead{(3)} & \colhead{(4)} & \colhead{(5)} & \colhead{(6)} & \colhead{(7)} &  
\colhead{(8)} & \colhead{(9)} & \colhead{(10)} & \colhead{(11)} & \colhead{(12)} & \colhead{(13)}
}
\startdata
BR~0331$-$1622    & 5.99 & 17.7 & 4.64 & 47.3 & $-$29.3 & {\phn}1.90$^{+0.88}_{-0.62}$ & 
{\phn}8.5$^{+4.0}_{-2.8}$ & {\phn}6.81 & 45.1 & 45.3 & $-1.86^{+0.09}_{-0.08}$ & $<2.9$\tablenotemark{i} \\
BR~0353$-$3820    & 1.51 & 17.9 & 3.86 & 47.2 & $-$29.2 & 11.30$^{+1.92}_{-1.68}$ & 
44.7$^{+7.6}_{-6.7}$      & 37.04      & 45.8 & 46.0 & $-$1.54$^{+0.07}_{-0.06}$ & 5.2\tablenotemark{i} \\ 
BR~0418$-$5723    & 1.72 & 17.8 & 4.24 & 47.3 & $-$29.2 & {\phn}1.76$^{+0.95}_{-0.66}$ & 
{\phn}7.0$^{+3.1}_{-2.6}$ & {\phn}5.69 & 45.0 & 45.2 & $-1.87^{+0.09}_{-0.10}$ & \nodata \\ 
BR~0424$-$2209    & 2.51 & 18.0 & 3.52 & 47.2 & $-$29.0 & {\phn}2.14$^{+0.92}_{-0.66}$ & 
{\phn}8.6$^{+3.7}_{-2.6}$ & {\phn}6.84 & 45.1 & 45.3 & $-$1.81$\pm{0.08}$ & $<3.8$\tablenotemark{i} \\
PSS~0747$+$4434   & 5.25 & 18.4 & 2.44 & 47.0 & $-$28.6 & {\phn}1.32$^{+0.79}_{-0.53}$ & 
{\phn}5.7$^{+3.5}_{-2.2}$ & {\phn}4.66 & 44.9 & 45.1 & $-$1.81$\pm{0.10}$ & $<1.5$\tablenotemark{j} \\
PSS~1058$+$1245   & 2.05 & 17.7 & 4.64 & 47.3 & $-$29.3 & {\phn}0.59$^{+0.58}_{-0.31}$ & 
{\phn}2.4$^{+2.3}_{-1.3}$ & {\phn}1.88 & 44.5 & 44.7 & $-2.07^{+0.13}_{-0.14}$ & $<0.8$\tablenotemark{j} \\
BR~1117$-$1329    & 5.05 & 18.3 & 2.67 & 47.0 & $-$28.5 & {\phn}0.43$^{+0.57}_{-0.28}$ & 
{\phn}1.8$^{+2.5}_{-1.2}$ & {\phn}1.37 & 44.3 & 44.5 & $-2.03^{+0.15}_{-0.18}$ & $<4.7$\tablenotemark{i} \\
PSS~1506$+$5220   & 1.68 & 18.3 & 2.67 & 47.0 & $-$28.6 & $<0.99$ & 
$<3.9$ & $<3.02$ & $<44.7$ & $<44.9$ & $<-1.90$ & $<1.4$\tablenotemark{j} \\
PSS~1646$+$5514   & 2.34 & 17.4 & 6.12 & 47.4 & $-$29.5 & {\phn}1.24$^{+0.74}_{-0.50}$ & 
{\phn}5.0$^{+3.0}_{-2.0}$ & {\phn}3.78 & 44.8 & 45.0 & $-$2.00$\pm{0.10}$ & $<0.1$\tablenotemark{k} \\ 
BR~2213$-$6729    & 2.73 & 18.5 & 2.22 & 47.0 & $-$28.6 & {\phn}3.66$^{+1.08}_{-0.86}$ & 
14.9$^{+4.4}_{-3.5}$ & 12.17 & 45.3 & 45.5 & $-$1.64$\pm{0.07}$ & \nodata \\ 
PSS~2344$+$0342   & 5.56 & 18.2 & 2.93 & 47.1 & $-$28.8 & $<0.59$ & 
$<2.6$ & $<2.05$ & $<44.5$ & $<44.7$ & $<-1.98$ & $<4.5$\tablenotemark{i} \\
\tableline
\enddata
\tablecomments{Luminosities are computed using 
$H_{0}$=70~km~s$^{-1}$~Mpc$^{-1}$, $\Omega_{\rm M}$=0.3, 
and $\Omega_{\Lambda}$=0.7.}
\tablenotetext{a}{From Dickey \& Lockman (1990) in units of 
$10^{20}$~cm$^{-2}$ .}
\tablenotetext{b}{Rest-frame 2500~\AA\ flux density in units of 
$10^{-27}$~erg~cm$^{-2}$~s$^{-1}$~Hz$^{-1}$.}
\tablenotetext{c}{Observed count rate computed in the 0.5--2~keV band 
in units of $10^{-3}$~counts~s$^{-1}$.}
\tablenotetext{d}{Galactic absorption-corrected flux in the observed 
0.5--2~keV band in units 
of $10^{-15}$~erg~cm$^{-2}$~s$^{-1}$. These fluxes and the following 
\xray\ parameters have been corrected for the ACIS quantum efficiency decay 
at low energy.} 
\tablenotetext{e}{Rest-frame 2~keV flux density in units of 
$10^{-32}$~erg~cm$^{-2}$~s$^{-1}$~Hz$^{-1}$.} 
\tablenotetext{f}{Rest-frame 2--10~keV luminosity in units of erg~s$^{-1}$.}
\tablenotetext{g}{Errors have been computed following the 
``numerical method'' described in $\S$~1.7.3 of Lyons (1991); 
both the statistical uncertainties on the \xray\ count rates and the effects 
of the observed ranges of the \xray\ and optical continuum shapes have been 
taken into account (see $\S$3 of V01 for details).} 
\tablenotetext{h}{Radio-loudness parameter, defined as 
$R$ = $f_{\rm 5~GHz}/f_{\rm 4400~\mbox{\scriptsize\AA}}$ (rest frame; e.g., 
Kellermann et al. 1989). 
The rest-frame 5~GHz flux density is computed from the observed 1.4~GHz 
flux density (mainly taken from the FIRST and the NVSS) assuming a radio 
power-law slope of $\alpha=-0.8$, with $f_{\nu}\propto~\nu^{\alpha}$. 
For two quasars, BR~0418$-$5723 and BR~2213$-$6729, no radio measurements 
are available.}
\tablenotetext{i}{1.4~GHz flux density from FIRST 
(Becker, White, \& Helfand 1995) .}
\tablenotetext{j}{1.4~GHz flux density from NVSS (Condon et al. 1998).}
\tablenotetext{k}{1.4~GHz flux density from Carilli et al. (2001).}
\label{tab3}
\end{deluxetable}
\end{turnpage}

\clearpage

\begin{deluxetable}{ccccccc}
\tablecolumns{7} 
\tabletypesize{\scriptsize}
\tablewidth{0pc} 
\tablecaption{Results from Joint Spectral Fitting}
\tablehead{
\colhead{Quasars used} & \colhead{Number of} & \colhead{} & \colhead{} & 
\colhead{\xray} & \colhead{} & \colhead{$N_{\rm H,z}$\tablenotemark{a}} \\
\colhead{in the fitting} & \colhead{Quasars} & \colhead{Median $z$} & \colhead{Median \mb} & 
\colhead{Counts} & \colhead{$\Gamma$} & \colhead{(cm$^{-2}$)} 
}
\startdata
All RQQs\tablenotemark{b}                      & 48 & 4.43 & $-$28.54 & 872 & 1.93$^{+0.10}_{-0.09}$ & $<4.95\times10^{21}$ \\
All but the 3 highest S/N QSOs                 & 45 & 4.43 & $-$28.48 & 674 & 1.92$\pm{0.10}$        & $<7.48\times10^{21}$ \\
Most optically luminous half of the sample     & 24 & 4.30 & $-$29.06 & 563 & 1.97$^{+0.13}_{-0.10}$ & $<6.50\times10^{21}$ \\
Least optically luminous half of the sample    & 24 & 4.60 & $-$28.25 & 309 & 1.88$^{+0.18}_{-0.15}$ & $<1.19\times10^{22}$ \\
Highest redshift half of the sample            & 24 & 4.83 & $-$28.10 & 373 & 1.98$^{+0.19}_{-0.16}$ & $<1.76\times10^{22}$ \\
Lowest redshift half of the sample             & 24 & 4.20 & $-$28.75 & 499 & 1.92$\pm{0.12}$        & $<4.86\times10^{21}$ \\
\tableline
\enddata
\label{tab4}
\tablenotetext{a}{Column density in the quasar rest frame.}
\tablenotetext{b}{The slightly different spectral-fit result reported in 
$\S$6 ($\Gamma=1.93\pm{0.09}$) is referred to the power-law model without absorption.}
\end{deluxetable}

\clearpage

\begin{figure}
\figurenum{1}
\centerline{\includegraphics[angle=0,width=\textwidth]{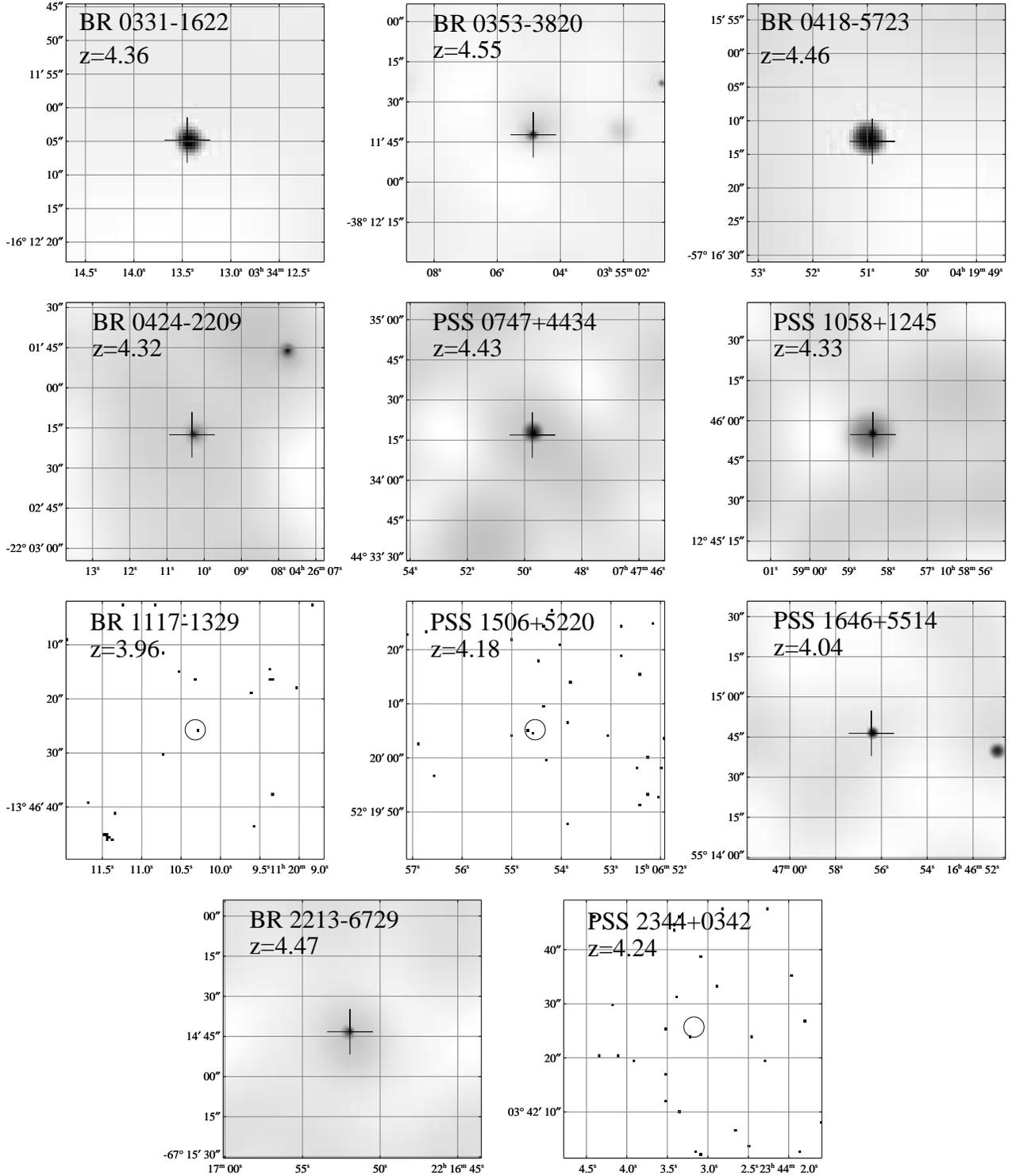}}
\vskip -3.0cm
\figcaption{\scriptsize
Full-band (0.5--8~keV) images of the quasars observed by \chandra\ and 
presented in this paper. The panels are 98\arcsec $\times$ 98\arcsec\ 
for all of the images, except for the source detected with two counts 
(BR~1117$-$1329; the two counts fall in the same pixel) and those without 
\xray\ detections (PSS~1506$+$5220 and PSS~2344$+$0342); for these three 
sources, the panels are 49\arcsec $\times$ 49\arcsec. 
North is up, and East to the left. 
The images of the sources with $>2$ counts have been adaptively smoothed at 
the 2$\sigma$ level; the optical positions of the quasars are marked 
by crosses. For the three sources with $\leq2$ counts, 
the raw (un-smoothed) images are shown; 
in these cases, \hbox{2\arcsec--radius} circles around the 
optical positions of the quasars are shown. We note that, although two 
photons fall within the \hbox{2\arcsec--radius} circle centered on 
the position of PSS~1506$+$5220, this is not considered a detection 
(see $\S$2.1).}
\end{figure}

\clearpage
\begin{figure}
\figurenum{2}
\centerline{\includegraphics[angle=-90,width=\textwidth]{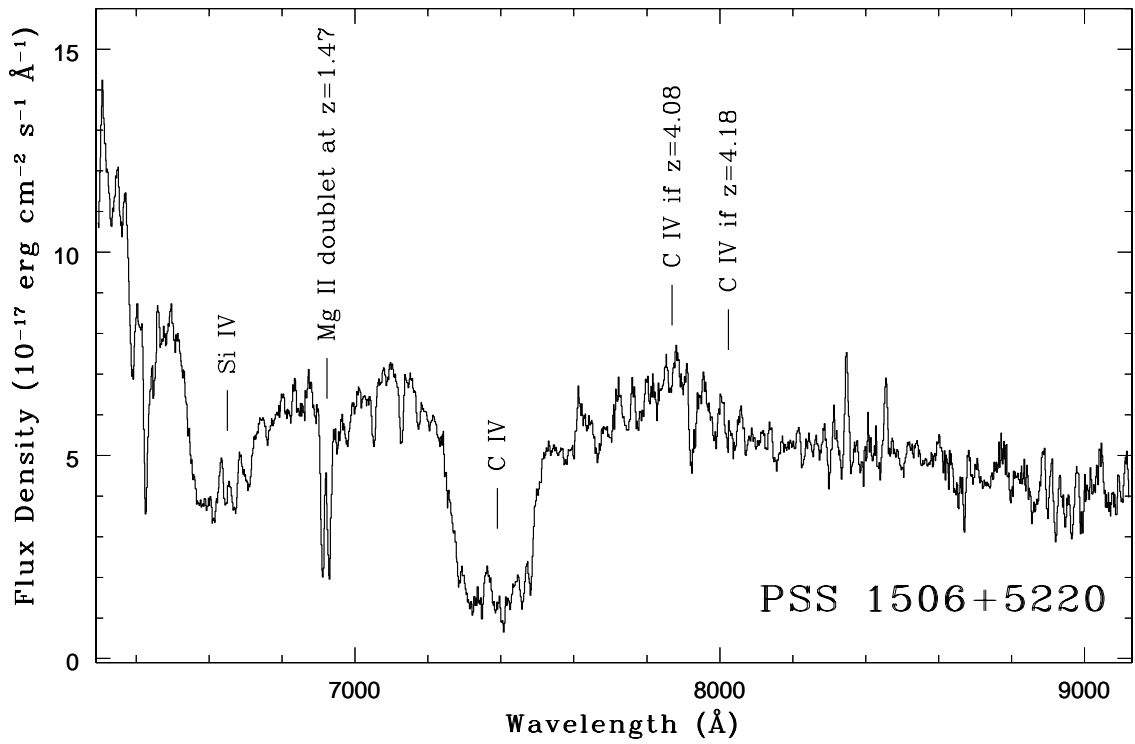}}
\figcaption{\footnotesize 
The spectrum of PSS~1506$+$5220 obtained with the Low Resolution Spectrograph 
of the Hobby-Eberly Telescope. The data have a spectral resolution of 
$\approx$~1100.  The two prominent broad absorption troughs by \ion{Si}{4} 
and \ion{C}{4} are labelled, 
as are the strong \ion{Mg}{2} absorption doublet at $z$=1.4711 
and the expected location of the \ion{C}{4} emission line at two redshifts 
(see $\S$3.2 for details).} 
\end{figure}

\clearpage

\begin{figure}
\figurenum{3}
\centerline{\includegraphics[angle=0,width=\textwidth]{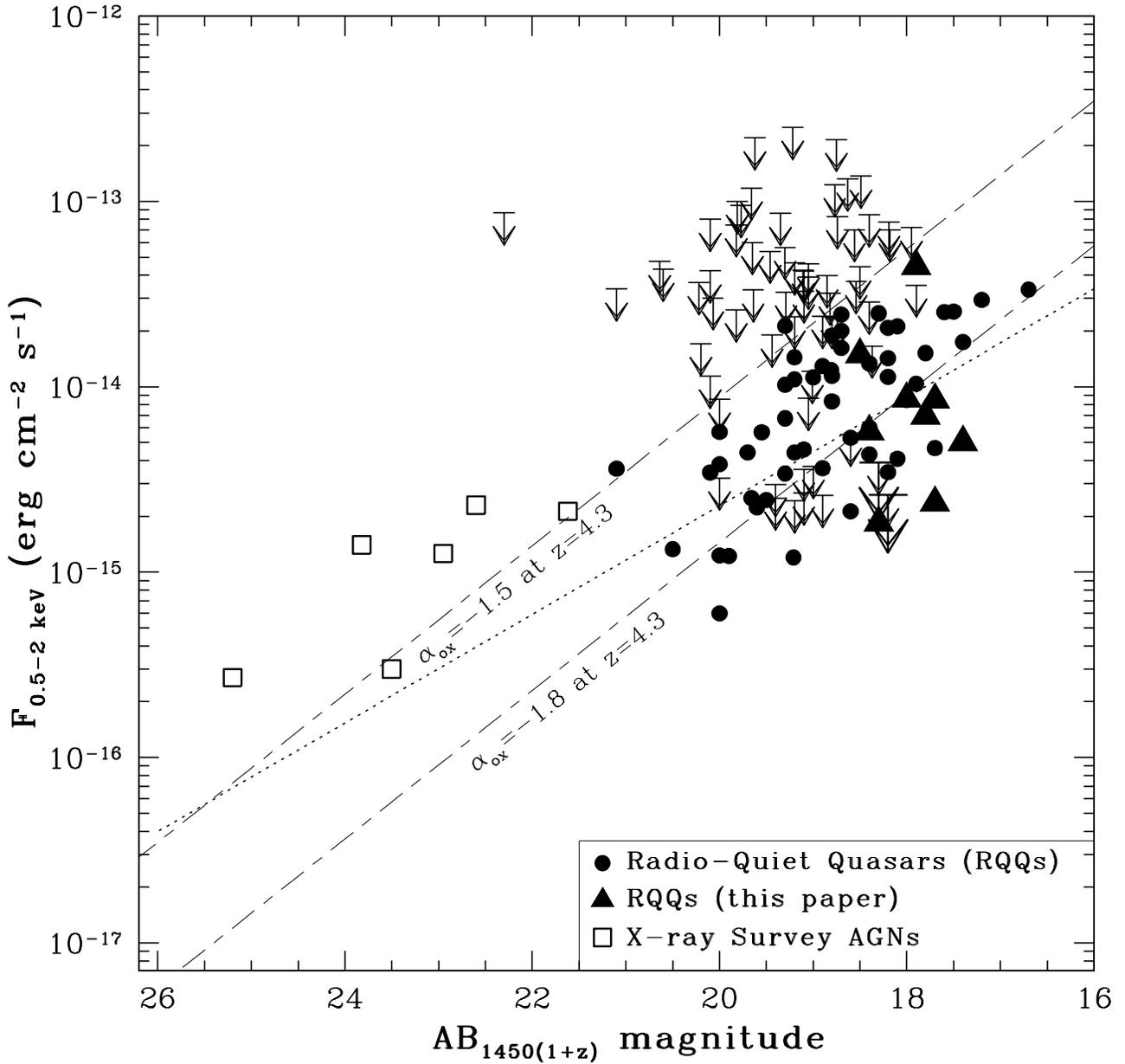}}
\figcaption{\footnotesize 
Observed-frame, Galactic absorption-corrected \hbox{0.5--2~keV} flux versus 
\ab1450\ magnitude for $z\ge$3.96 radio-quiet AGN, mainly quasars. 
The objects presented in this paper are shown as large filled triangles 
(\xray\ detections) and large downward-pointing arrows (upper limits); 
RQQs from previous \xray\ observations (Kaspi, Brandt, \& Schneider 2000; 
V01; Brandt et al. 2002; Bechtold et al. 2003; V03a; V03b) 
are shown as filled circles (\xray\ detections) and small downward-pointing 
arrows (upper limits). 
AGN detected by \xray\ surveys (Schneider et al. 1998; Silverman et al. 2002; 
V02; Castander et al. 2003; Treister et al. 2004) 
are shown as open squares. 
The slanted dashed lines show the \aox=$-$1.5 and \aox=$-$1.8 loci at $z$=4.3 
(the average redshift of the present sample); 
the dotted line shows the best-fit correlation reported in $\S$5 for the 
sample of high-redshift, optically selected RQQs observed by \chandra.}
\end{figure}

\clearpage

\begin{figure}
\figurenum{4}
\centerline{\includegraphics[angle=0,width=\textwidth]{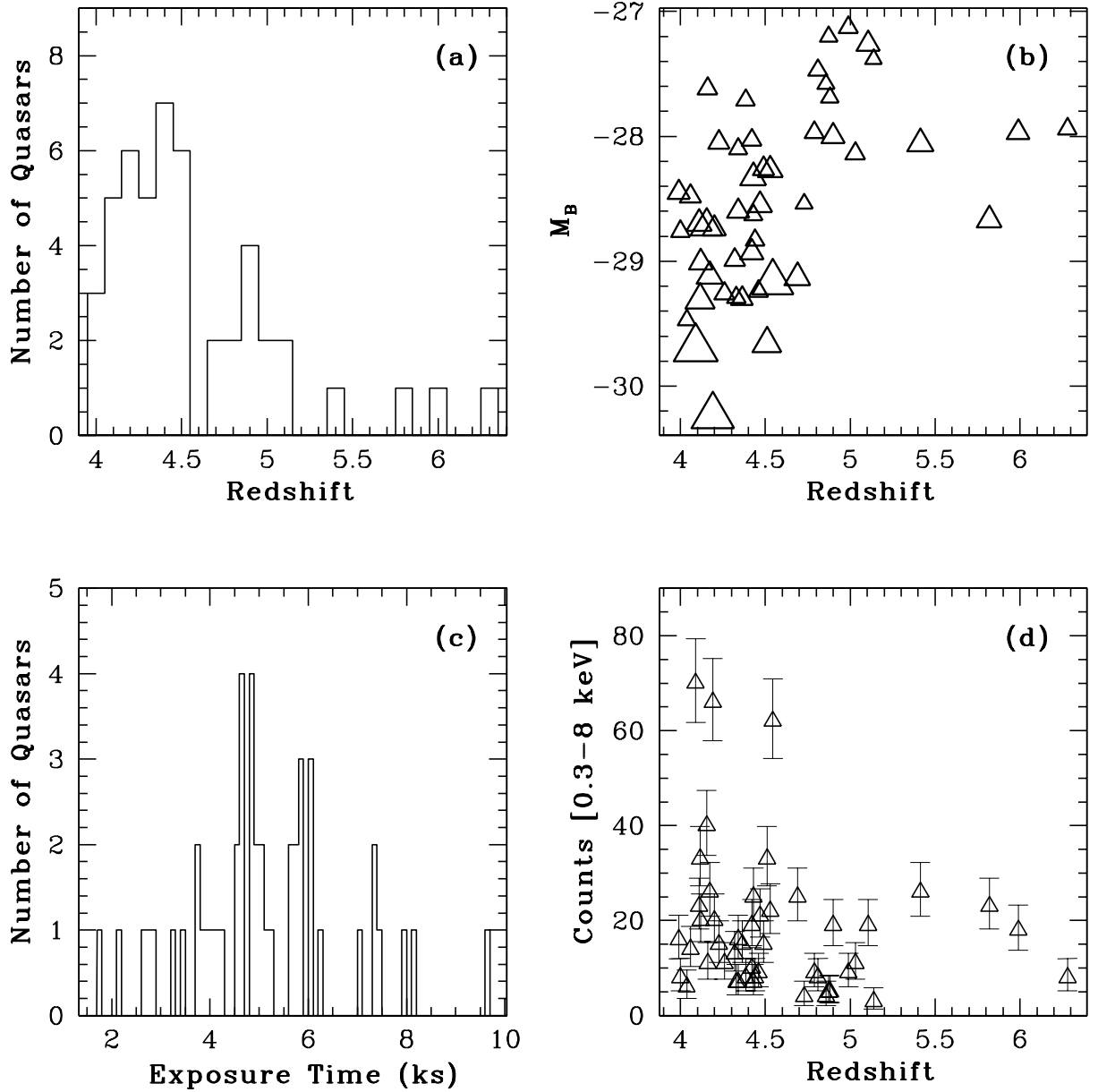}}
\figcaption{\footnotesize Relevant information for the 48 quasars used in the 
\xray\ spectral analysis: 
(a) histogram of redshift; 
(b) redshift vs. absolute $B$-band magnitude (the sizes of the symbols 
increase with the number of counts); 
(c) histogram of the exposure times (corrected for detector dead time); 
(d) redshift vs. number of counts and relative errors (computed following 
Gehrels 1986).}
\end{figure}

\clearpage

\begin{figure}
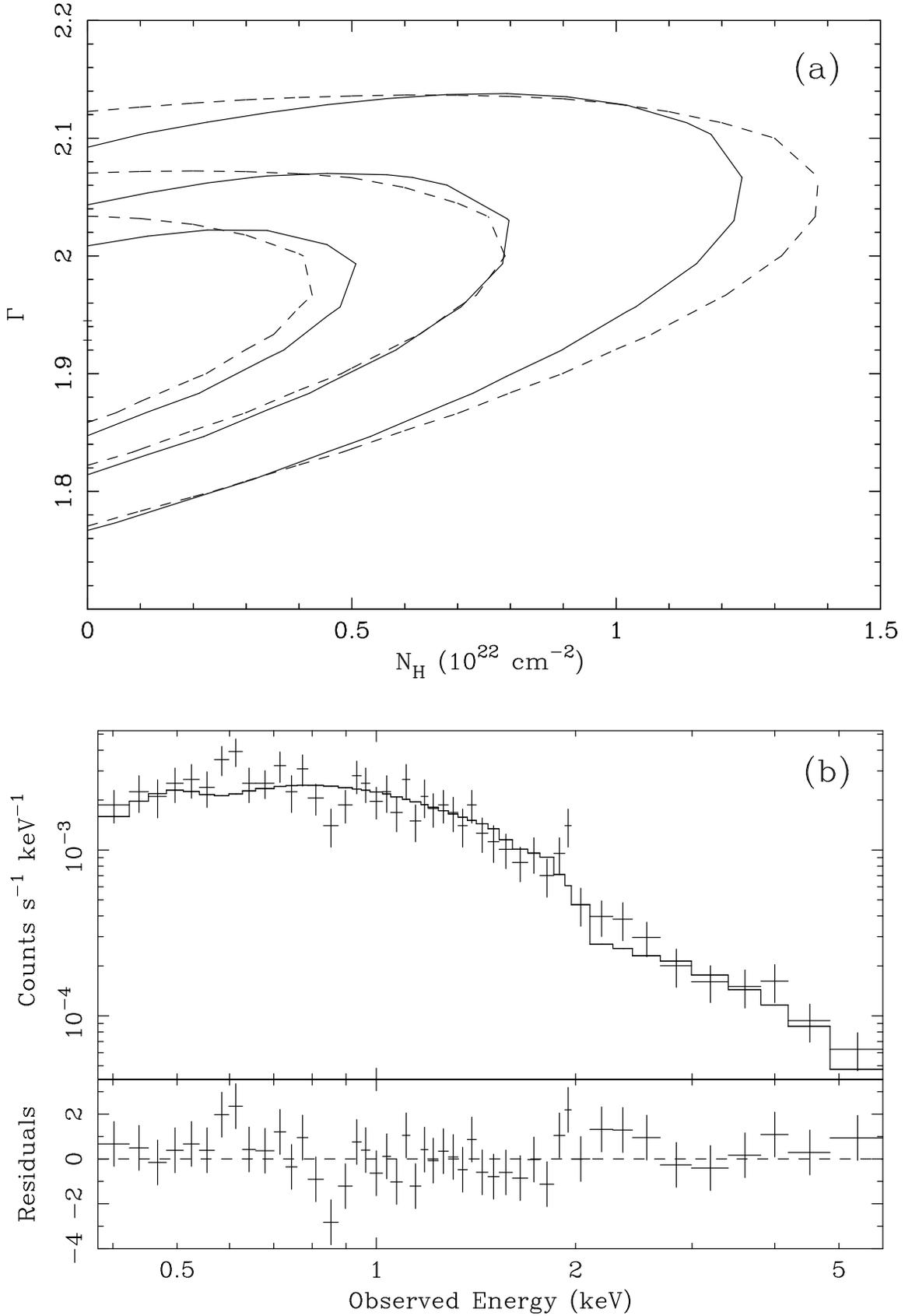

\figurenum{5}
\centerline{\includegraphics[angle=-90,width=0.80\textwidth]{fig5a.ps}}
\vglue1.5cm\hglue-0.5cm
\centerline{\includegraphics[angle=-90,width=0.83\textwidth]{fig5b.ps}}
\vglue0.4cm
\figcaption{\scriptsize
(a) 68, 90, and 99\% confidence regions for the photon index vs. 
intrinsic column density derived from joint \xray\ spectral fitting of 
the 48 RQQs at $z\ge$3.99 detected by \chandra\ with more than 2 counts. 
The average redshift of the sample is $z$=4.57, while its median redshift 
is $z$=4.43. 
The solid contours have been obtained using all of the source counts, 
while the dashed contours show the results obtained in the rest-frame 
\hbox{$\approx$~2.2--40~keV} band common to all of the quasars of 
the present sample. 
(b) Combined spectrum of the 48 RQQs at $z\ge$3.99 fitted with a power law 
and Galactic absorption (top panel) and data-to-model residuals 
in units of $\sigma$ (bottom panel).} 
\end{figure}

\clearpage

\begin{figure}
\figurenum{6}
\centerline{\includegraphics[angle=0,width=0.48\textwidth]{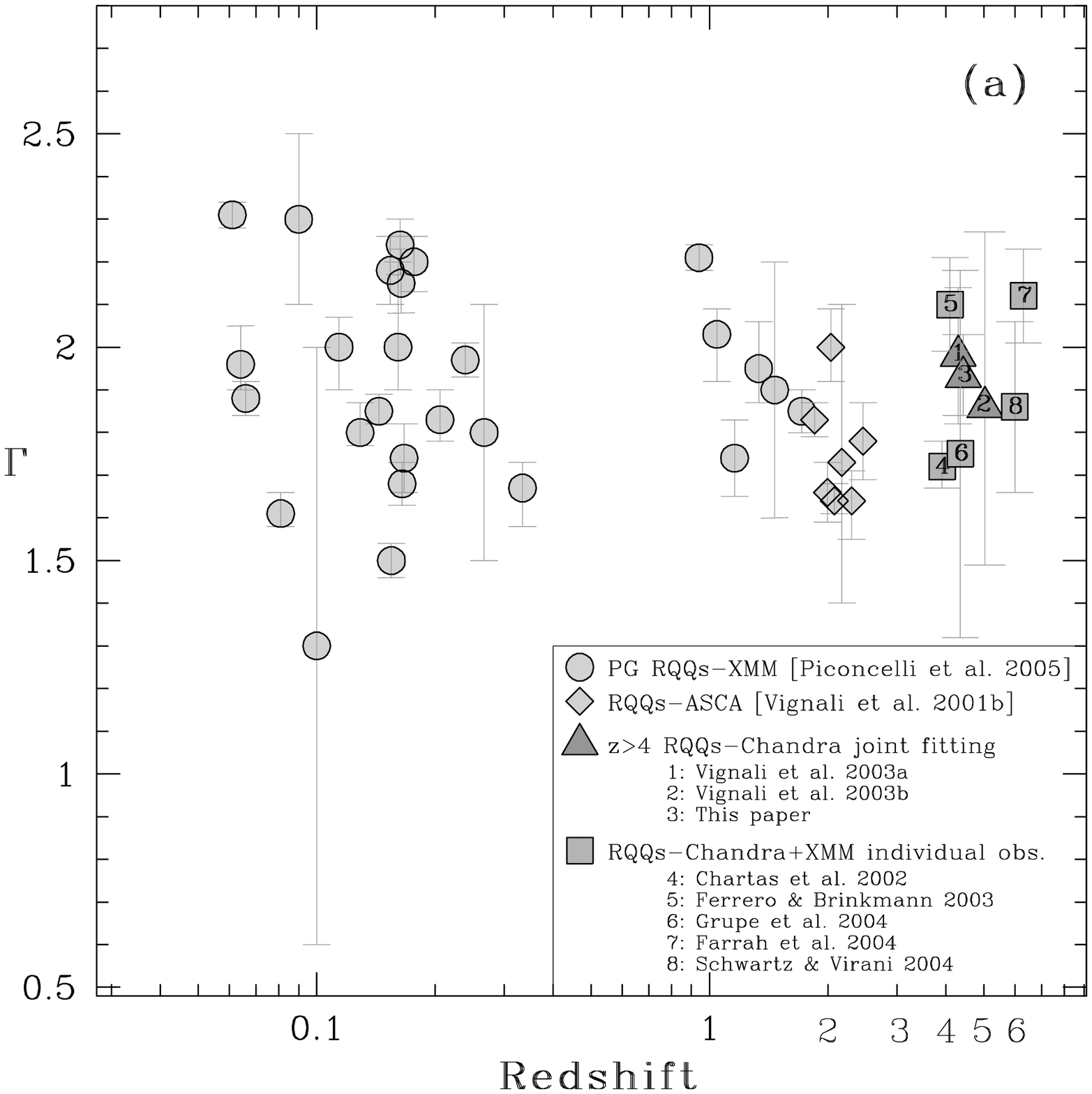}}
\centerline{\includegraphics[angle=0,width=0.48\textwidth]{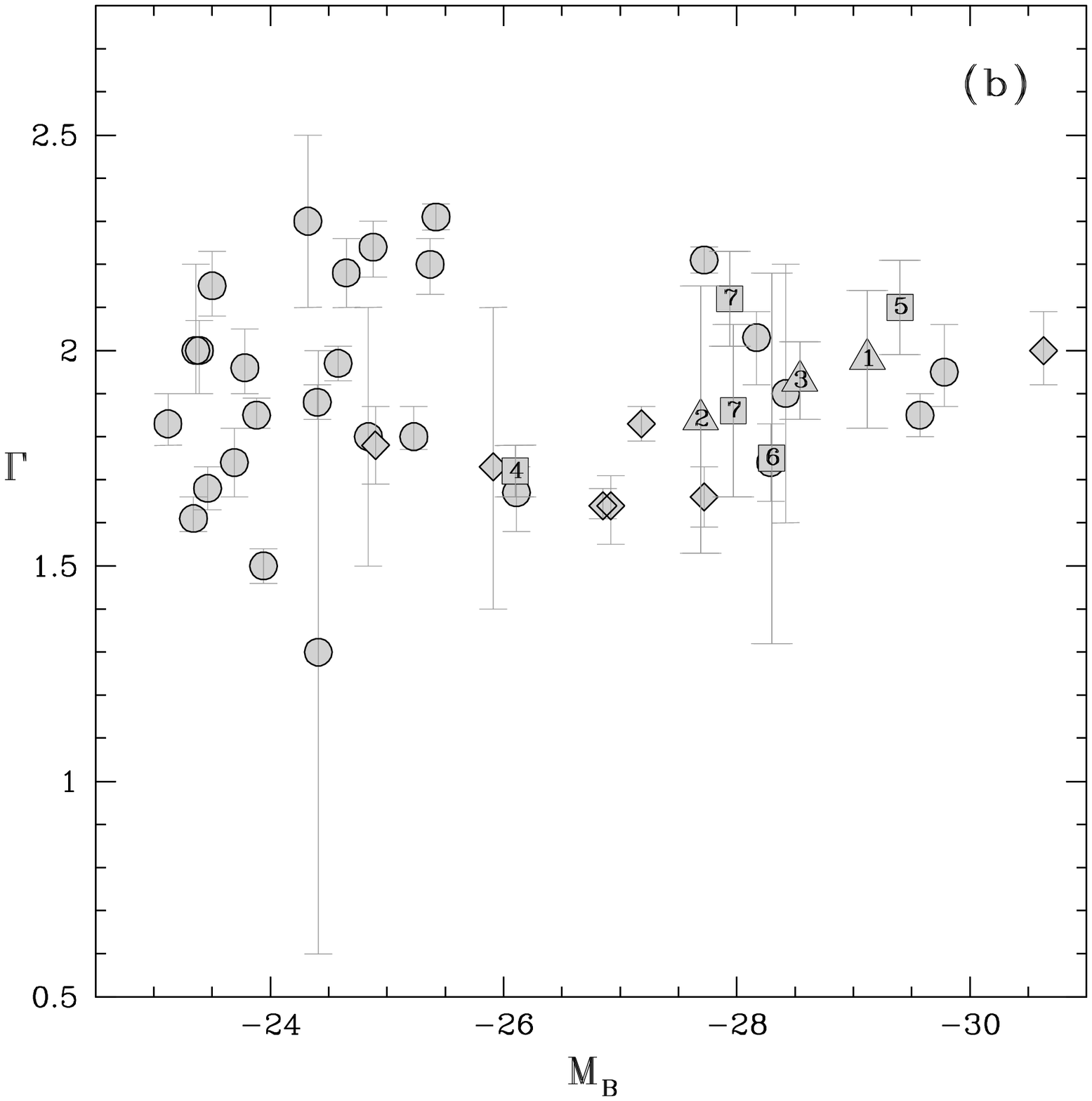}}
\figcaption{\footnotesize 
(a) Hard \xray\ photon index vs. redshift for a compilation of RQQs. 
The circles at $z<1.8$ indicate RQQs from the Bright Quasar Survey (PG) 
analyzed by Piconcelli et al. (2005), while the diamonds 
indicate the $z\approx1.8-2.5$ RQQs from Vignali et al. (2001b) sample. 
At higher redshift, the data points indicate joint spectral fitting 
(triangles) and single-object spectroscopy 
(squares); numbers in the data points correspond to 
the citations given in the figure legend. 
(b) Hard \xray\ photon index vs. \mb\ for the same quasars.}
\end{figure}

\clearpage

\begin{figure}
\figurenum{7}
\centerline{\includegraphics[angle=0,width=\textwidth]{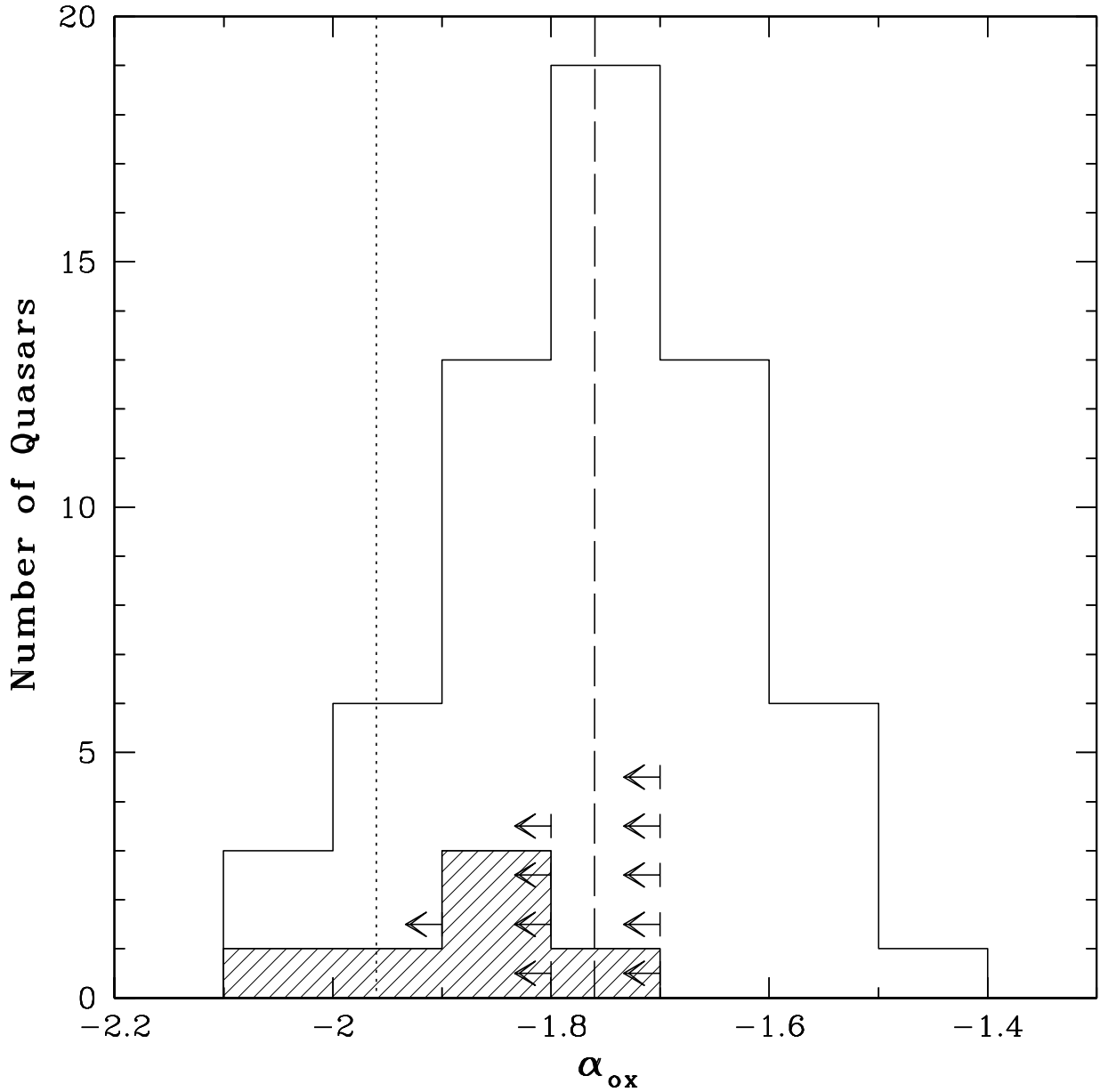}}
\figcaption{\footnotesize 
\aox\ distribution for the optically selected RQQs observed by \chandra\ 
at $z\geq3.96$; note that here we have included also the two-count source 
BR~1117$-$1329 at $z$=3.96 which was not used in the joint \xray\ spectral 
fitting. 
The shaded area indicates the BALQSOs; the areas with arrows indicate the 
95\% confidence upper limits. 
The long-dashed (dotted) line indicates the average \aox\ (using {\sc asurv}) 
for high-redshift non-BAL RQQs (BALQSOs) observed by \chandra. 
The typical uncertainty in the \aox\ measurements is $\approx$~0.15.}
\end{figure}

\clearpage

\appendix

\section{New \rosat\ upper limits for $z>$4 SDSS DR2 quasars}
Here we report some additional \xray\ flux upper limits for the 
Sloan Digital Sky Survey Data Release~2 (SDSS DR2; Abazajian et al. 2004) 
quasars at $z>4$. These were obtained with \rosat\ PSPC and HRI 
using only the inner $\approx$~20\arcmin\ of these instruments. 
The 3$\sigma$ upper limits (see Table~A1) have been derived in a similar way 
to those shown in Table~4 of V01 using the {\sc sosta} command in the 
{\sc ximage} package (Giommi et al. 1992) Version 4.2. 
The \ab1450\ magnitudes have been derived using the method described in 
$\S$2.2 of VBS03. 

\begin{deluxetable}{lcccccc}
\tablecolumns{7}
\tabletypesize{\footnotesize}
\tablewidth{0pt}
\tablecaption{Properties of $z>4$ SDSS DR2 Quasars in \rosat\ Fields}
\tablehead{ 
\colhead{Object} & \colhead{$z$} & \colhead{$AB_{1450(1+z)}$} & 
\colhead{$M_B$} & \colhead{$f_{\rm x}$\tablenotemark{a}} & 
\colhead{$\alpha_{\rm ox}$} & \colhead{$R$\tablenotemark{b}} 
}
\startdata
SDSS~004054.65$-$091526.79 & 4.973 & 19.3 & $-$27.6 & $<3.24$ & $<-1.34$ & $<5.6$   \\
SDSS~100645.60$+$462717.25 & 4.440 & 20.1 & $-$26.7 & $<3.02$ & $<-1.29$ & $<136.7$ \\
SDSS~101053.00$+$531144.82 & 4.506 & 20.1 & $-$26.7 & $<1.14$ & $<-1.45$ & $<10.1$  \\
SDSS~123735.47$+$642936.00 & 4.334 & 20.1 & $-$26.6 & $<4.22$ & $<-1.23$ & $<32.7$  \\
SDSS~140146.53$+$024434.72 & 4.375 & 18.7 & $-$28.0 & $<2.36$ & $<-1.54$ & $<2.8$   \\
SDSS~150730.63$+$553710.83 & 4.499 & 20.2 & $-$26.6 & $<1.71$ & $<-1.36$ & $<11.1$  \\
SDSS~213243.26$+$010633.91 & 4.032 & 20.2 & $-$26.4 & $<3.65$ & $<-1.25$ & $<10.5$  \\
\tableline
\enddata
\tablenotetext{a}{Galactic absorption-corrected flux in the observed 
\hbox{0.5--2~keV} band in units of \hbox{$10^{-14}$~erg~cm$^{-2}$~s$^{-1}$.}} 
\tablenotetext{b}{Radio-loudness parameter.}
\label{A1}
\end{deluxetable}

\vfill\eject

\section{Optical, X-ray, and Radio Properties of $z\ge4$ Quasars in Archival \chandra\ Observations}

Here we report the main optical, X-ray, and radio properties for $z\ge4$ quasars with 
unpublished \chandra\ observations. Note that BR~1202$-$0725 was detected also by \rosat\ 
(Kaspi, Brandt, \& Schneider 2000).

\vglue -3cm
\begin{deluxetable}{lccccccccccc}
\tablecolumns{12}
\tabletypesize{\tiny}
\tablewidth{0pt}
\tablecaption{Optical, X-ray, and Radio Properties of $z\ge4$ Quasars in Archival \chandra\ Observations}
\tablehead{ 
\colhead{Object} & \colhead{$z$} & \colhead{\ab1450} & \colhead{$f_{\rm 2500~\AA}$} & 
\colhead{$\log (L_{\rm 2500~\AA})$\tablenotemark{a}} & \colhead{$M_B$} & 
\colhead{$f_{\rm 0.5-2~keV}$} & \colhead{$f_{\rm 2~keV}$} & 
\colhead{$\log (L_{\rm 2~keV})$\tablenotemark{b}} & \colhead{$\log (L_{\rm 2-10~keV})$} & 
\colhead{\aox} & \colhead{$R$}
}
\startdata
%
PSS~0134$+$3307  & 4.53 & 18.8 &  1.69 & 31.80 & $-$28.3 &      11.5   & 9.45    & 27.55    & 45.44    & $-$1.63  & $<8.1$ \\
PSS~0808$+$5215  & 4.44 & 18.2 &  2.93 & 32.03 & $-$28.8 & {\phn}3.5   & 2.81    & 27.01    & 44.90    & $-1.93$  & 0.2    \\
BR~~1202$-$0725   & 4.70 & 18.0 &  3.52 & 32.14 & $-$29.1 & {\phn}8.5   & 7.21    & 27.45    & 45.34    & $-1.80$  & $<1.3$ \\
BR~~1600$+$0724   & 4.38 & 19.3 &  1.06 & 31.58 & $-$27.7 & {\phn}6.8   & 5.43    & 27.28    & 45.21    & $-1.65$  & $<3.4$ \\
BR~~2235$-$0301\tablenotemark{c}
                 & 4.25 & 18.9 &  1.54 & 31.72 & $-$28.1 &      $<2.6$ & $<2.03$ & $<26.84$ & $<44.73$ & $<-1.87$ & $<8.5$ \\
PSS~2322$+$1944  & 4.12 & 17.9 &  3.86 & 32.09 & $-$29.0 &      10.4   & 7.92    & 27.41    & 45.30    & $-1.80$  & 0.2    \\
\tableline
\enddata
\tablecomments{Units are the same as in Table~3.} 
\tablenotetext{a}{Rest-frame 2500~\AA\ luminosity density (\lumh).}
\tablenotetext{b}{Rest-frame 2~keV luminosity density (\lumh).}
\tablenotetext{c}{BALQSO.}
\label{B1}
\end{deluxetable}

\end{document}